\documentclass{article}
\usepackage{spconf,amsmath,graphicx}
\usepackage{amssymb,amsmath,epsfig}
\usepackage{latexsym}
\usepackage{graphicx,multirow,threeparttable,float,balance,array,bm,subcaption,caption}
\usepackage{cite}
\usepackage{hyperref}
\usepackage[english]{babel}

\title{Optimization of Speaker Extraction Neural Network with Magnitude and Temporal Spectrum Approximation Loss}
\name{Chenglin Xu$^{1,2}$, Wei Rao$^3$, Eng Siong Chng$^{1,2}$, Haizhou Li$^{2,3}$\thanks{Wei Rao contributed to this work before joining National University of Singapore. This research is supported by the National Research Foundation Singapore under its AI Singapore Programme (and other co-funders, where applicable). [AISG-100E-2018-006]. Accepted in ICASSP 2019.}}
\address{
  $^1$ School of Computer Science and Engineering, Nanyang Technological University, Singapore \\
  $^2$ Temasek Laboratories@NTU, Nanyang Technological University, Singapore \\
  $^3$ Department of Electrical and Computer Engineering, National University of Singapore, Singapore\\
}
\begin{document}
%
\maketitle
\begin{abstract}
The SpeakerBeam-FE (SBF) method is proposed for speaker extraction. It attempts to overcome the problem of unknown number of speakers in an audio recording during source separation. The mask approximation loss of SBF is sub-optimal, which doesn't calculate direct signal reconstruction error and consider the speech context. To address these problems, this paper proposes a magnitude and temporal spectrum approximation loss to estimate a phase sensitive mask for the target speaker with the speaker characteristics. Moreover, this paper explores a concatenation framework instead of the context adaptive deep neural network in the SBF method to encode a speaker embedding into the mask estimation network. Experimental results under open evaluation condition show that the proposed method achieves 70.4\% and 17.7\% relative improvement over the SBF baseline on signal-to-distortion ratio (SDR) and perceptual evaluation of speech quality (PESQ), respectively. A further analysis demonstrates 69.1\% and 72.3\% relative SDR improvements obtained by the proposed method for different and same gender mixtures.
\end{abstract}
\vspace{-5pt}
\section{Introduction}\vspace{-5pt}
\label{sec:intro}
Speech is the most natural way of human-machine interface. However, the performance of speech recognition is significantly affected in face of background noise and interference speech \cite{watanabe2017new,xiao2016study}. To naturally interact with machines, we hope that machine can give selective attention to the speech of interest and ignore the rest with speech separation or speaker extraction techniques.

Recent deep learning based methods, such as Deep Clustering (DC) \cite{hershey2016deep,isik2016single,wang2018alternative}, Deep Attractor Network (DANet) \cite{chen2017deep}, Permutation Invariant Training (PIT) methods \cite{yu2017permutation,kolbaek2017multitalker,xu2018single,xu2018shifted}, have significantly advanced the performance of multi-taker speech separation. However, the number of speaker has to be known in prior. The PIT methods need this prior information during training. Although it's not necessary for the DC and DANet methods in the training stage, it's required during inference to form the clusters equal to the number of speakers in the mixture. 

Since the number of speaker information is always unknown in practice, the usefulness of speech separation technique is greatly limited. To address this limitation, two research directions have been explored. The first one is to iteratively reconstruct the speech for every speaker \cite{kinoshita2018listening}. The iteration procedure is terminated by either a stop-flag or a threshold of the residual mask. The other one is target speaker extraction that only extract the target speaker's voice from a mixture given the speaker information \cite{delcroix2018single,wang2018deep}. This technique is practical to the applications where only registered speakers need to be responded, for example, the speaker verification application \cite{rao2019target}. The SpeakerBeam-FE (SBF) method \cite{delcroix2018single} exploited a context adaptive deep neural network (CADNN) \cite{delcroix2016context} to track the target speaker with a speaker adaptation layer. A mask approximation loss is calculated between the estimated mask and ideal binary mask as the objective to train the network. However, the mask approximation loss is sub-optimal for two reasons. One is that the loss function is not directly targeting reconstruction errors. The other is that the speech context is not considered in the loss function.

To address these problems, this paper proposes a magnitude and temporal spectrum approximation loss to estimate a phase sensitive mask for the target speaker with the speaker characteristics. The loss is calculated on magnitude and its dynamic information (i.e., delta and acceleration) to constrain the extracted speech for temporal continuity. In addition, we explore a concatenation framework to encode the speaker embedding into the mask estimation instead of the adaptation structure in the SBF method. The same joint optimization strategy is applied to the auxiliary network and mask estimation network. The auxiliary network encodes the speaker information to speaker embeddings, which are repeatedly concatenated with the activations to estimate masks with the proposed loss. 

Section \ref{sec:problem} describes the speaker extraction problem with a mask. The details of the proposed magnitude and temporal spectrum approximation loss and concatenation framework are discussed in Section \ref{sec:system}. Section \ref{sec:exp} reports the experimental setup and results. The conclusions are given in Section \ref{sec:con}.
\vspace{-5pt}
\section{Speaker Extraction with A Mask}\vspace{-5pt}
\label{sec:problem}
The target speech extraction aims to extract the target speaker's voice $x(n)$ from a linearly mixed single channel microphone signal $y(n)$ given a different speech segment $a(n)$ of the target speaker. The mixed signal is,
\begin{equation} \label{eq:discret_signal}
y[n] = x[n] + \sum_{i=1}^I z_i[n]
\end{equation}
where $z_i[n]$ might be any number of interference speech or background noise.

Given the mixed signal $y(n)$ and enroll signal $a(n)$, the goal is to estimate $\hat{x}[n]$ that is close to $x[n]$. This problem has been formulated as a supervised learning task. It estimates a filter (i.e., mask) for the target speaker with the supervised information ( i.e., ideal binary mask (IBM) in \cite{delcroix2018single}). Previous works with magnitude spectrum approximation loss have shown better performance than the mask approximation loss calculated between the IBM and estimated mask in speech enhancement \cite{wang2014training,erdogan2015phase} and separation \cite{yu2017permutation,xu2018single}.

The extracted magnitude $|\hat{X}(t,f)|$ of the target speaker is obtained by
\begin{equation}
|\hat{X}(t,f)| = M(t,f) \odot |Y(t,f)|
\end{equation}
where $\odot$ indicates element-wise multiply. $M(t,f)$ is the estimated phase sensitive mask (PSM) and $|Y(t,f)|$ is the spectrogram representation of mixed signal. The phase of the mixed speech $\angle Y(t,f)$ is used to reconstruct the time domain signal $\hat{x}[n]$.

\section{The Speaker Extraction Network}\vspace{-5pt}
\label{sec:system}

\begin{figure}[tb]
\begin{center}
\includegraphics[width=80mm]{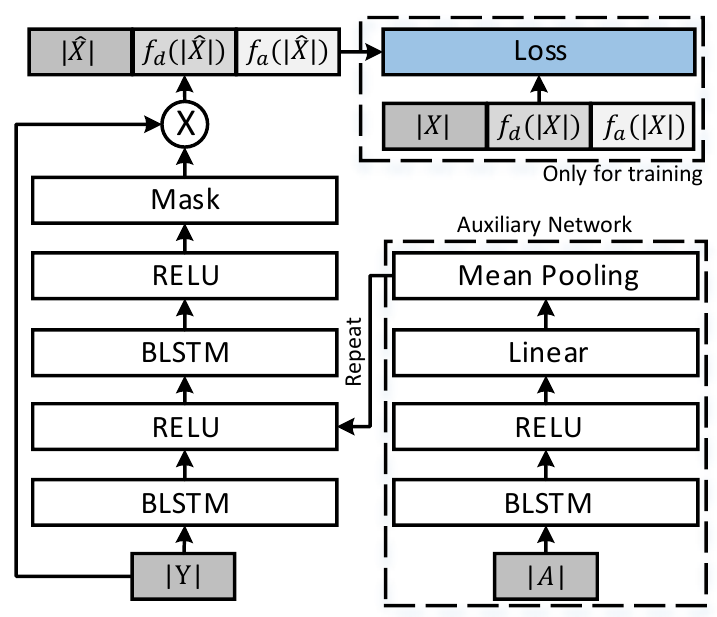} \vspace{-5pt}
\caption{The concatenation framework for monaural target speaker extraction with magnitude and temporal spectrum approximation loss. During inference, the upper dotted box is not necessary. And the system takes mixture ($|Y|$) and a target speaker's voice ($|A|$) in and outputs the extraction ($|\hat{X}|$). $f_d(\cdot)$ and $f_a(\cdot)$ are delta and acceleration function.}
\label{fig:system}
\end{center}
\vspace{-23pt}
\end{figure}

The SBF method \cite{delcroix2018single} uses an auxiliary network to estimate adaptation weights, which weight the sub-layers in the adaptation layer for mask estimation with a mask approximation loss. The loss is defined as,
\begin{equation} \label{eq:obj_mask}
    J_1 = \frac{1}{T}\sum ||M - M_{ibm}||_F^2
\end{equation}
However, the mask approximation loss is sub-optimal, because the mask approximation loss doesn't directly reflect the signal reconstruction error between the estimated signal and the true signal. In addition, the loss between the estimated and ideal masks is computed individually without considering the context information. The continuity characteristics of speech should be used to constrain the mask estimation for temporal continuity.

\subsection{Magnitude and Temporal Spectrum Approximation Loss}
\label{subsec:loss}

To address the limitations in mask approximation loss defined in Eq. \ref{eq:obj_mask}, we propose to compute a magnitude and temporal spectrum approximation loss (MTSAL) to replace the mask approximation loss (MAL) in SBF method. We name it as the SBF-MTSAL approach. The loss is firstly calculating signal reconstruction error between the extracted magnitude and clean magnitude with phase difference. It also computes the errors across the dynamic information (i.e., delta and acceleration) based on the extracted magnitude and clean magnitude for temporal continuity. The proposed magnitude and temporal spectrum approximation loss is defined as, 
\begin{equation} \label{eq:obj_mtsa}
    \begin{aligned}
J_2 &=\frac{1}{T}\sum (||M\odot |Y|-|X|\odot cos(\theta_y-\theta_x)||_F^2 \\
&+ w_d||f_d(M\odot |Y|)-f_d(|X|\odot cos(\theta_y-\theta_x))||_F^2 \\
&+ w_a||f_a(M\odot |Y|)-f_a(|X|\odot cos(\theta_y-\theta_x))||_F^2)
\end{aligned}
\end{equation}
where $M$ is the estimated phase sensitive mask for target speaker. $|Y|$ and $|X|$ are the magnitudes of the mixture and the target speaker's clean speech. $\theta_y$ and $\theta_x$ are phase angles of the mixture and target speaker's clean speech. $w_d$ and $w_a$ are the weights (set as 4.5 and 10.0 in this work). $f_d(\cdot)$ and $f_a(\cdot)$ are the function to compute delta and acceleration. Since the acceleration computation function is computing the delta twice, we only given the delta computation function \cite{furui1986speaker} as,
\begin{equation}
    f_d(v(t))=\frac{\sum_{l=1}^Ll\times(v(t+l)-v(t-l))}{\sum_{l=1}^L2l^2}
\end{equation}
where $v(t)$ is a time frame of magnitude. $L$ is the contextual window and is set to 2.

\subsection{The Concatenation Framework}
\label{subsec:concat}
To learn speaker-dependent transforms for the acoustic features, the i-vector is concatenated with acoustic features as inputs for speaker adaptation of acoustic model in speech recognition \cite{saon2013speaker,senior2014improving,karanasou2014adaptation}. Inspired by this idea, 
this paper explores a concatenation framework to learn transforms for the target speaker. The framework repeatedly concatenates the speaker embeddings from an auxiliary network with the mixture representations in a mask estimation network. The proposed direct signal reconstruction error of MTSAL is used to estimate a phase sensitive mask for the target speaker based on the concatenated representations. We name it as SBF-MTSAL-Concat, as shown in Fig. \ref{fig:system}.

Different from the auxiliary network in the SBF baseline using frame level features to obtain speaker information, our auxiliary network encodes the speaker information into an embedding by a BLSTM to use contextual information from history and future based on a whole utterance. It takes the magnitude ($|A|\in R^{T\times F}$) of the target speaker's different utterance in and outputs a $D$ dimensional embedding vector ($V\in R^{D}$) using a mean pooling over all frames.

\section{Experiments and Discussion}\vspace{-5pt}
\label{sec:exp}

\subsection{Data}
\label{subsec:data}
We generated a two speakers mixture database\footnote{The database simulation code is available at: \url{https://github.com/xuchenglin28/speaker_extraction}} without background noise and reverberation at sampling rate of 8kHz based on the WSJ0 corpus \cite{garofolo1993csr}. The simulated database with average length of 6s was divided into three sets: training set ($20,000$ utterances), development set ($5,000$ utterances), and test set ($3,000$ utterances). Specifically, the utterances from $50$ male and $51$ female speakers in the WSJ0 ``si\_tr\_s'' set were randomly selected to generate the training and development set at various SNR uniformly chosen between 0dB and 5dB. Similarly, the test set was created by randomly mixing the utterances from $10$ male and $8$ female speakers in the WSJ0 ``si\_dt\_05'' set and ``si\_et\_05'' set. Since the speakers in the test set were different from the training and development sets, the test set was used to evaluate the speaker independent performance and regarded as open condition (OC) evaluation. The development set was considered as closed condition (CC) to tune parameters due to the same speaker used in training and development sets.

In the two-speaker mixture simulation, the first selected speaker was chosen as target speaker, the other one was interference speaker. The utterance of the target speaker from the original WSJ0 corpus was used as reference speech. Another utterance of this target speaker, which was different from the reference speech, was randomly selected to be used as input to the auxiliary network to obtain target speaker information.

\subsection{Experimental Setup}
\label{subsec:setup}
A short-time Fourier transform (STFT) was used with a window length of $32$ms and a shift of $16$ms to obtain the magnitude features from both the input mixture for mask estimation network and input target speech for auxiliary network. The normalized square root hamming window was used. The same magnitude features were used in all experiments.

\vspace{-5pt}
\subsubsection{The SBF baseline}
\vspace{-5pt}
In the auxiliary network, $2$ feed-forward hidden layers with relu activation function were built on the input layer and had $512$ nodes in each layer. A following feed-forward linear layer with $30$ nodes, which were equal to the number of the sub-layers in the adaptation layer of the mask estimation network, computed the adaptation weights. The weights were obtained by averaging these $30$ dimensional outputs over all the frames. In the mask estimation network, the BLSTM following the input layer had $512$ cells in forward and backward directions, respectively. The following adaptation layer had $30$ sub-layers. Each sub-layer had $512$ nodes with $1024$ dimensional inputs from the outputs of previous BLSTM. The $30$ dimensional weights from the auxiliary network were used to weight these sub-layers, respectively. After that, the activations of all the sub-layers were summed as the output of the adaptation layer. Another $2$ feed-forward hidden layers with relu activation function were used. Each hidden layer had $512$ nodes. The output layer had $129$ nodes to predict the mask for the target speaker. The mask approximation loss defined in Eq. \ref{eq:obj_mask} was used to optimize the network.

\vspace{-5pt}
\subsubsection{The SBF-MTSAL}
\vspace{-5pt}
The network configuration is the same as SBF baseline. The only difference is the proposed MTSAL defined in Eq.~\ref{eq:obj_mtsa} was used to optimize the network.

\vspace{-5pt}
\subsubsection{The SBF-MTSAL-Concat}
\vspace{-5pt}
In the auxiliary network, we fed magnitude of a target speaker's utterance to a BLSTM with $256$ cells in each forward and backward direction. A following feed-forward hidden layer with relu activation function had $256$ nodes. Then a linear layer with $30$ nodes was used to obtain a $30$ dimensional speaker embedding with a mean pooling over all frames\footnote{Although the dimension of the speaker embedding can be tuned to different number without parameter explosion problem in the SBF-MTSAL-Concat method, we keep it same as the dimension of the adaptation weights in the baseline \cite{delcroix2018single}.}. In the mask estimation network, a BLSTM with $512$ cells in both directions was built on top of the input mixture. The speaker embedding was repeatedly concatenated to each frame of the activation from the BLSTM. Then the concatenations were fed to a relu hidden layer, a BLSTM and another relu hidden layer. The number of node or cell was set to $512$. The mask layer output a $129$ dimensional mask, which was element-wise multiplied with the input mixture to extract the target speaker.

In all experiments, the learning rate started from $0.0005$ and scaled down by $0.7$ when the training loss increased on the development set. The minibatch size was set to $16$. The network was trained with minimum $30$ epochs and stopped when the relative loss reduction was lower than $0.01$. The Adam algorithm \cite{kingma2014adam} was used to optimize the network. We evaluated the performance with the criteria of signal-to-distortion ratio (SDR) \cite{vincent2006performance} and perceptual evaluation of speech quality (PESQ) \cite{rix2001perceptual}.

\subsection{Results}
\label{subsec:results}
Table \ref{tbl:methods_comp} summarizes the SDR and PESQ performances of the SBF baseline \cite{delcroix2018single} and our proposed techniques. Compared with the original mixture, the PESQ performance of the SBF baseline reported in Table~\ref{tbl:methods_comp} even degrades under closed condition. Because the mask approximation loss is not direct signal reconstruction error. The lower error between the estimated and ideal masks doesn't mean better speech quality. The estimated mask may harm the speech context by forcing it to be close to ideal binary mask with mask approximation loss. When applying the proposed magnitude and temporal spectrum approximation loss, results of SBF-MTSAL in Table~\ref{tbl:methods_comp} shows that the SDR and PESQ performance are relatively improved by 53.5\% and 14.7\%, respectively, comparing with the SBF baseline under open condition. In the closed condition, the relative SDR and PESQ improvements are 59.9\% and 17.0\%, individually. It shows that the magnitude and temporal information in the objective function could constrain the extracted speech for temporal continuity. Since the speakers are unseen during training in open condition, the proposed approach can extend well to unseen speakers compared with the improvements in closed condition. It's a speaker independent system.

\begin{table}[ht]
\centering \caption{ SDR (dB) and PESQ in a comparative study of different techniques under closed condition (CC) and open condition (OC). ``SBF-MTSAL'' is for the magnitude and temporal spectrum approximation loss (Eq. \ref{eq:obj_mtsa}) in SBF method instead of mask approximation loss (Eq. \ref{eq:obj_mask}). ``SBF-MTSAL-Concat'' is the concatenation framework with magnitude and temporal spectrum approximation loss instead of the adaptation structure in SBF method. ``Paras'' means the number of model parameters.} 
\centerline{
\footnotesize
\begin{tabular}{|r|*{5}{c|}}
\hline
\multirow{2}{*}{Method} & \multirow{2}{*}{Paras} & \multicolumn{2}{c|}{CC} & \multicolumn{2}{c|}{OC} \\ \cline{3-6}
 & & SDR & PESQ & SDR  & PESQ  \\
\hline
\hline
Mixture & - & 2.60 & 2.32 & 2.60 & 2.31 \\
\hline
SBF \cite{delcroix2018single} & 19.3M & 6.48 & 2.30 & 6.45 & 2.32 \\ 
SBF-MTSAL & 19.3M & 10.36 & 2.69 & 9.90 & 2.66 \\ 
SBF-MTSAL-Concat & 8.9M & \textbf{11.39} & \textbf{2.77} & \textbf{10.99} & \textbf{2.73} \\ \hline
\end{tabular}} 
\label{tbl:methods_comp}
\end{table}

Different from the auxiliary network using a DNN in the SBF-MTSAL approach, we observed that the performance is further improved by the concatenation framework with a BLSTM to learn speaker embedding in the auxiliary network using its history and future information. By applying the concatenation framework, the number of parameters is significantly reduced. And it improves the SDR from 9.90dB to 10.99dB with a 11.0\% relative improvement in open condition. Meanwhile, the PESQ has a 2.6\% relative improvement. Compared with the SBF baseline, the SBF-MTSAL-Concat method achieves a 70.4\% and a 17.7\% relative SDR and PESQ improvements under open condition.

The performance for different and same gender mixture is further analyzed and summarized in Table \ref{tbl:gender}. The performance on different gender mixture is always better than the same gender, since the characteristics between same gender speakers are closer and more difficult to discriminate than the different gender. Experimental results in Table \ref{tbl:gender} demonstrate that the proposed MTSAL and concatenation framework significantly advance the performance of the SBF baseline. Specifically, SBF-MTSAL-Concat could achieve 69.1\% and 72.3\% relative improvements over the SBF baseline on SDR for different and same gender conditions. This also agrees with the conclusions in Table~\ref{tbl:methods_comp}.

\begin{table}[t] 
\centering \caption{SDR (dB) and PESQ in a comparative study of different and same gender mixture under open condition.} 
\centerline{
\small
\begin{tabular}{|r|*{4}{c|}}
\hline
\multirow{2}{*}{Method} & \multicolumn{2}{c|}{SDR} & \multicolumn{2}{c|}{PESQ} \\ \cline{2-5}
 & Diff. & Same & Diff.  & Same  \\
\hline
\hline
Mixture & 2.51 & 2.69 & 2.29 & 2.34 \\
\hline
SBF \cite{delcroix2018single} & 7.61 & 5.13 & 2.42 & 2.19 \\ 
SBF-MTSAL & 12.27 & 7.17 & 2.85 & 2.44 \\ 
SBF-MTSAL-Concat & \textbf{12.87} & \textbf{8.84} & \textbf{2.90} & \textbf{2.54} \\ \hline
\end{tabular}} 
\label{tbl:gender} 
\end{table}

\section{Conclusions}
\label{sec:con}
In this paper, we proposed a magnitude and temporal spectrum approximation loss for speaker extraction network. Besides this, a concatenation framework is explored to encode the target speaker information into a mask estimation network instead of adaptation technique in SBF method. Experimental results show that the proposed loss with the concatenation framework can significantly improve the SDR and PESQ performance and reduce the number of parameters to be half. A further analysis shows that the relative improvement for the same gender condition is higher than the different gender condition.



\bibliographystyle{IEEEbib}
\bibliography{refs}

\end{document}